# The seasonality of air ticket prices before and after the pandemic

Alessandro V. M. Oliveira✈

Aeronautics Institute of Technology, São José dos Campos, Brazil

✈Corresponding author. Technological Institute of Aeronautics. Praça Marechal Eduardo Gomes, 50. 12.280-250 - São José dos Campos, SP - Brazil.
E-mail address: alessandro@ita.br .

*Abstract*: **This study investigates price seasonality in the Brazilian air transport industry, emphasizing the impact of the COVID-19 pandemic on domestic airline pricing strategies. Given potential shifts in demand patterns following the global health crisis, this study explores possible long-term structural changes in the seasonality of Brazilian airfare. We analyze an open dataset of domestic city pairs from 2013 to 2023, employing an econometric model developed using Stata software. Our findings indicate alterations in seasonal patterns and long-term trends in the post-pandemic era. These changes underscore potential shifts in the composition of leisure and business travelers, along with the cost pressures faced by airlines.**

*Keywords*: air transport, airlines, econometrics.

*Reproducibility*: The data, code, and final study used in this study are available on open-access online platforms.

## I. INTRODUCTION

Price seasonality describes how the prices of goods and services fluctuate throughout the year due to variations in supply and demand associated with seasons or specific periods. Factors such as weather patterns, holidays, events, income, and consumption habits can directly impact product availability and consumer spending. Businesses and consumers need to understand price seasonality patterns to make informed purchasing, selling, and marketing decisions, adapt to market trends, and maximize value.

Seasonal fluctuations are a crucial characteristic of the air transport industry, affecting airlines, airports, air traffic management, and passengers alike. Airlines use their understanding of seasonality to strategically plan flight schedules, capacity, and fares. This allows them to maximize profits during high-demand periods (like holidays) and mitigate losses during off-seasons. Airports and air traffic control use these insights to anticipate demand, ensuring efficient operations and resource allocation. For consumers, awareness of price seasonality can help them find cost-effective travel options during lower-demand periods. Seasonality also drives competition, motivating airlines to innovate their offerings and services to remain attractive throughout the year.

The COVID-19 pandemic profoundly impacted the air transport industry, potentially altering seasonal patterns due to changes in supply chains and consumer behavior. Shifts in work environments, like remote work, can modify corporate travel demand. Consumer preferences may also change, with travelers favoring local destinations over international ones.

This study examines price seasonality in Brazilian air transport, focusing on how COVID-19 influenced pricing strategies within the domestic market. Our goal is to determine if demand shifts following the easing of travel restrictions have altered the annual pricing cycle for air travel within Brazil. We'll analyze pre- and post-pandemic price data to uncover trends and variations, providing insights into how Brazilian airlines have reacted to this unprecedented challenge. This study draws on open data spanning over a decade of domestic flight pricing dynamics and ultimately informs an econometric model of air transport pricing.

This study is structured as follows. Following the Introduction, Section II ("Literature Review") examines relevant previous studies and their findings. Section III ("Empirical Modeling") details the methodology employed. This includes variable generation, data analysis, econometric model construction, and the interpretation of results – all undertaken to explore the pandemic's impact on seasonal pricing within domestic routes. Finally, Section IV ("Conclusions") provides a summary of key findings, discusses implications for managers and policymakers, and outlines potential areas for future research.

## II. LITERATURE REVIEW

The literature on seasonality in air transport and tourism examines various aspects, considering not only cyclical patterns of tourist demand but also strategic airline behavior and airport profitability.

Halpern (2011) explores seasonal variations in passenger demand at Spanish airports. While this variation is generally low, it differs significantly between airports. The research indicates that tourist destination airports exhibit greater seasonality than those in large metropolitan areas with a balanced mix of leisure and business travelers unless the tourist destination enjoys consistent visitation year-round. Additionally, the study finds no direct correlation between airport size and seasonality. Detailed analysis of Ibiza Airport demonstrates that international routes and charter flights lead to increased seasonality, while domestic and regular flights tend to lessen it – except low-cost regular flights, which can also heighten seasonality. Póvoa and Oliveira (2008) employ econometric modeling to assess the impact of holidays on airfares. The authors focus on three-day holiday events, using a database of fares collected online between 2008 and 2010 for Brazil's largest city, São Paulo. Another study on Brazilian airline pricing is by Varella, Frazão, and Oliveira (2017), which uses proxies for seasonal effects among the covariates of a regression model. In addition to dummies for holiday dates, the authors use a set of dummy variables for the day of the week and month of flight purchase and departure to control for other factors. In contrast, Brito, Oliveira, and Dresner (2021) implement a more general approach to control for seasonality, using temporal fixed effects in a Brazilian airline pricing model. For a general discussion on air ticket prices in Brazil, see Resende & Oliveira (2023).

Cazanova, Ward, and Holland (2014) investigate the persistence of tourism habits, specifically concerning flights to Florida. They explore how specific events and seasonal factors influence consumers' travel decisions. Their analysis delves into the tendency of tourists to return to Florida, highlighting drivers beyond merely economic factors. Holidays, significant events like the Daytona 500, and adverse weather conditions all play a key role. The research emphasizes the pronounced seasonality of Florida's tourism, with a peak in March coinciding with spring break and a decline in September as classes resume. Additionally, the findings suggest hurricanes and fires have a surprisingly low impact on domestic air travel demand. This points to Florida's resilience as a tourist destination and underscores the importance of understanding travel patterns for effective marketing and destination management. Zuidberg (2017) examines the link between airport seasonality and profitability, establishing an optimal seasonality level for maximizing profits. The study demonstrates an inverted "U"-shaped relationship, indicating that profitability suffers under both extremely high and extremely low seasonality. This is due to fluctuating operational costs and air traffic challenges. The finding stresses the complexities involved in managing airport seasonality, emphasizing the need for a balanced approach in pursuit of operational efficiency and financial success.

Merkert and Webber (2018) analyze how airlines manage seasonal demand fluctuations with strategic fare adjustments and capacity management. Their model, tested with data from airlines across the globe, suggests that companies should prioritize maximizing occupancy (seat sales) over fare increases to optimize profit during peak seasons. The authors observe that this strategy is often not implemented in reality, highlighting a potential area for improved revenue management practices within the airline industry. Claussen, Essling, and Peukert (2018) study how a company's strategic flexibility influences its entry into markets where demand fluctuates, with a focus on the US airline industry. Their findings demonstrate that routes with high demand variability generally attract fewer companies. However, firms with greater strategic flexibility show an increased willingness to compete in these uncertain markets. This effect is amplified when demand becomes even less predictable, underscoring the importance of adaptability when operating in volatile environments. Turrión-Prats and Duro (2018) create a framework to evaluate how seasonality influences tourism in Spain and apply it to explore underlying market patterns. Their findings reveal that since 2008, the concentration of tourism demand within specific peak periods has increased significantly, despite overall growth in the sector. Analysis suggests that market prices, exchange rates, and crucially, the income levels of tourists play a substantial role in shaping these seasonal trends. Finally, Ridderstaat and Croes (2020) propose a model to assess tourism seasonality, specifically distinguishing between changes that happen across different seasons and variations occurring within a single season.

Research by Zou, Reynolds-Feighan, and Yu (2022) reveals a correlation between high seasonality and reduced fleet utilization for Low-Cost Carriers (LCCs). They suggest that geographical route diversification could lessen this adverse effect. This finding has strategic importance for LCCs aiming to maximize fleet utilization while contending with fluctuating seasonal demand.

Dobruszkes, Decroly, and Suau-Sanchez (2022) report that over one-third of airports worldwide experience pronounced seasonality, a factor less evident in larger airports. They stress the significance of considering airport size, climate, and local geography when analyzing seasonal fluctuations in air traffic. This underscores the multifaceted nature of airport management and the demand for adaptable strategies.

## III. EMPIRICAL MODELING

The empirical analysis in this study utilizes a database housed in the data repository of the Air Transport Economics Center (NECTAR-ITA) on the Harvard Dataverse portal. This database, titled AVDATA-BR-CP, aggregates Brazilian aviation data into domestic city pairs (CP) on a monthly basis. The National Civil Aviation Agency (ANAC) serves as the primary data source. A detailed description of the data and variables used can be found in Oliveira (2024).

The following analysis employs commands from the Stata software language, though routines can be adapted to other languages. The accompanying Stata do-file is available for download (Oliveira, 2024a). To begin, we import the data using the following command:

```
. import delimited
  https://dataverse.harvard.edu/api/access/datafile/avd
  atabr_cp_cae/8171526, case(preserve) clear
```

We present the results of econometric analyses and modeling conducted on a sample extracted from a comprehensive database of imported data. Our sample focuses on domestic city pairs with sufficient demand density for an average aircraft size to exceed 50 seats. To ensure robust analysis, we only selected city pairs with at least five years (60 months) of observations recorded between 2013 and 2023 (through November 2023). The final panel comprises 487 city pairs and 131 periods, totaling 55,950 observations.

The following variables from the original AVSTATS-BR-CP database were selected to be used in the empirical model: price, km_great_circle_distance, jetfuel_price_org, nr_revenue_pax, market_concentration_hhi, and pc_load_factor. These variables were then adapted and renamed for inclusion in this study. The final econometric modeling employs the following variables derived from the original data: AirFare (average round-trip airline ticket price; source: ANAC). FuelPrice (regional aviation fuel price charged by the producer at the flight's origin; source: Brazilian National Agency for Petroleum, Natural Gas and Biofuels). PaxDens (paid passengers on direct flights between cities; source: ANAC); MktConc (city-pair market concentration index, calculated from ticket sales data; source: ANAC); LoadFactor (average aircraft utilization percentage on direct flights within the city pair); Pandemic (pandemic event dummy variable, indicating the period between February 2020, the first recorded COVID-19 case, and April 2022, when the Ministry of Health declared the end of the Public Health Emergency of National Importance due to COVID-19); Trend (five-year trend variable). All monetary variables have been deflated to reflect January 2024 values. Continuous variables were transformed into logarithms. Descriptive statistics of the variables are presented in Table 1.

**Table 1– Descriptive statistics of the variables**

| Variable   | N     | Mean | SD   | Min  | Max   |
|------------|-------|------|------|------|-------|
| AirFare    | 55950 | 6.25 | 0.40 | 4.11 | 8.29  |
| Distance   | 55950 | 6.65 | 0.68 | 4.68 | 7.95  |
| FuelPrice  | 55950 | 1.10 | 0.28 | 0.43 | 1.82  |
| PaxDens    | 55950 | 8.63 | 1.47 | 2.20 | 12.98 |
| MktConc    | 55950 | 8.61 | 0.39 | 7.80 | 9.21  |
| LoadFactor | 55950 | 4.33 | 0.15 | 2.57 | 4.61  |
| Pandemic   | 55950 | 0.18 | 0.39 | 0.00 | 1.00  |
| Trend      | 55950 | 1.10 | 0.62 | 0.02 | 2.18  |

To create seasonality variables, the following commands were first used:

```
. gen WintBreak = (Month==7)
. gen SummBrSearch = (Month==8 | Month==9 | Month==10 | Month==11)
. gen SummBreak = (Month==12 | Month==1 | Month==2)
. gen LowSeason = (Month==4 | Month==5 | Month==6)
```

Therefore, four dummies of periods of the year were created: WintBreak (month of July, which has school holidays), SummBrSearch (months between August and November, marked by intense price research for travel in the high summer season; it is a period of many business trips due to the greater activity of the economy typical of the second semester), SummBreak (high summer season), and LowSeason (low season, after the end of summer). Note that March serves as the base case for these dummies (and is therefore omitted). We employed a linear regression model with Multiple Fixed Effects (MFE) to conduct the regressions. The Stata command "refghdfe" (Correia, 2017) was used. In our analysis, we utilize city-pair fixed effects (CityPair). Since only one dimension is involved, this approach is equivalent to the traditional Fixed Effects estimator. The following commands were used to execute the regressions:

```
. reghdfe AirFare FuelPrice PaxDens MktConc LoadFactor Pandemic Trend, absorb(CityPair)
. est store WithoutSeas
. reghdfe AirFare FuelPrice PaxDens MktConc LoadFactor Pandemic Trend WintBreak SummBrSearch SummBreak LowSeason, absorb(CityPair)
. est store WithSeas
```

The results of the estimations without and with seasonality controls are found in Table 2. To generate the table, we used the "esttab" routine, present in the "estout" package, written by user (Jann, 2004):

```
. esttab WithoutSeas WithSeas, nocons nose not nogaps noobs b(%9.4f) varwidth(14) brackets aic(%9.0fc) bic(%9.0fc) ar2 scalar(N) sfmt(%9.0fc) addnote("Notes: Fixed Effect estimation")
```

**Table 2 – Models Without and With Seasonality Controls**

|              | (1) AirFare | (2) AirFare |
|--------------|-------------|-------------|
| FuelPrice    | 0.2931***   | 0.2897***   |
| PaxDens      | -0.0888***  | -0.0937***  |
| MktConc      | 0.0854***   | 0.0859***   |
| LoadFactor   | 0.3121***   | 0.2935***   |
| Pandemic     | -0.1674***  | -0.1690***  |
| Trend        | 0.0215***   | 0.0180***   |
| WintBreak    |             | -0.0009     |
| SummBrSearch |             | 0.0593***   |
| SummBreak    |             | 0.0145***   |
| LowSeason    |             | -0.0452***  |
| adj. R-sq    | 0.579       | 0.588       |
| AIC          | 8,539       | 7,265       |
| BIC          | 8,601       | 7,363       |
| N            | 55,950      | 55,950      |

Notes: Fixed Effect estimation
\* $p<0.05$, \*\* $p<0.01$, \*\*\* $p<0.001$

The findings in Table 2 align with expectations and are intuitively understandable. Our observations indicate that fuel prices (FuelPrice), market concentration (MktConc), and the utilization factor (LoadFactor) drive an increase in average airfare. Conversely, traffic volume (PaxDens) acts as a reducing factor, allowing airlines to leverage economies of scale at the route level (density economies) and incorporate those efficiencies into ticket pricing. Regarding the pandemic, our analysis suggests a 16-17% reduction in average passenger fares when all other factors are held constant – notably traffic density, which experienced a sharp decline. This estimated price decrease could be a sign of survival strategies adopted by airlines during the pandemic. Additionally, we see a long-term upward price trend (Trend) of approximately 2% every five years. All of these results hold statistical significance. Finally, Table 1 sheds light on the focal variables of this study – price seasonality. As expected, both the peak summer season (SummBreak) and its preceding search period (SummBrSearch) exhibit above-average fares. On the flip side, other periods show below-average costs, most notably the LowSeason when discounted tickets are most common. Notably, while the winter school break (WintBreak) wasn't statistically significant, it suggests pricing below that of peak summer periods.

Another key takeaway is the substantial ceteris paribus price increase (nearly 6%) during the research period before the high season – surpassing even the high season itself. This may imply limited intra-station summer availabilities, as choice seats on prime flights likely sell out in advance. Consequently, many tickets bought then are actually for the subsequent season. The presence of openings at less desirable times, along with the approaching end of peak season and the start of the low season, could contribute to the less pronounced price increases. Importantly, remember that these are ceteris paribus effects – calculated while holding all other factors constant.

Next, we outline a simple experiment that complements the one that generated the results shown in Table 2. This time, we model seasonality with greater detail. We aim to analyze the price increase cycle, beginning several months before the high season, extending through it, and concluding with the drop in the low season. To achieve this, we have replaced the grouped seasonal dummy variables (SummBrSearch, SummBreak, and LowSeason) with monthly equivalents employing "before" ("bef") and "after" ("aft") indicators to mark time relative to the beginning of each phase. For instance, SummBreak_bef4 denotes the fourth month before the summer starts (August), SummBreak_bef3 denotes September, and so forth. Likewise, LowSeason_aft1 signifies April (with March as the base case). The purpose of this granular dummy approach is to provide a dynamic view of price fluctuations throughout the year while holding all other factors constant. Here's the Stata code used to generate these dummies::

```
. gen SummBreak_bef4 = (Month==8)
. gen SummBreak_bef3 = (Month==9)
. gen SummBreak_bef2 = (Month==10)
. gen SummBreak_bef1 = (Month==11)
. gen SummBreak_aft0 = (Month==12)
. gen SummBreak_aft1 = (Month==1)
. gen SummBreak_aft2 = (Month==2)
. gen LowSeason_aft1 = (Month==4)
. gen LowSeason_aft2 = (Month==5)
. gen LowSeason_aft3 = (Month==6)
```

In this new experiment, we again use the MFE/Fixed Effects estimator. Below is the code used to generate the regressions and display the results:

```
. reghdfe AirFare FuelPrice PaxDens MktConc LoadFactor Pandemic Trend WintBreak SummBreak_* LowSeason_*, absorb(CityPair)
. est store GranularSeas
. esttab GranularSeas, nocons nose not nogaps noobs b(%9.4f) varwidth(14) brackets aic(%9.0fc)
```

```
bic(%9.0fc) ar2 scalar(N) sfmt(%9.0fc)
addnote("Notes: Fixed Effect estimation ")
```

Table 3 displays the results obtained.

**Table 3– Model with Granular Seasonality**

|                 | (1) AirFare |
|-----------------|-------------|
| FuelPrice       | 0.2910***   |
| PaxDens         | -0.0941***  |
| MktConc         | 0.0865***   |
| LoadFactor      | 0.2950***   |
| Pandemic        | -0.1677***  |
| Trend           | 0.0167***   |
| WintBreak       | -0.0008     |
| SummBreak_bef4  | 0.0563***   |
| SummBreak_bef3  | 0.0688***   |
| SummBreak_bef2  | 0.0812***   |
| SummBreak_bef1  | 0.0315***   |
| SummBreak_aft0  | 0.0609***   |
| SummBreak_aft1  | -0.0156**   |
| SummBreak_aft2  | 0.0020      |
| LowSeason_aft1  | -0.0492***  |
| LowSeason_aft2  | -0.0380***  |
| LowSeason_aft3  | -0.0482***  |
| adj. R-sq       | 0.591       |
| AIC             | 6,957       |
| BIC             | 7,117       |
| N               | 55,950      |

Notes: Fixed Effect estimation
* $p<0.05$, ** $p<0.01$, *** $p<0.001$

The results of the regressions are then displayed in a coefficient graph, generated using the "coefplot" routine (Jann, 2013). This approach facilitates a direct comparative analysis of the impact of seasonality coefficients across the specified periods. The coefplot generator command is as follows:

```
. coefplot GranularSeas, keep(WintBreak SummBreak_*
  SummBreak_* LowSeason_*) xline(0, lcolor(green)
  lpattern(dash)) scheme(s2color) level(95)
  recast(connected) lpattern(longdash) lwidth(0.1)
```

Figure 1 presents the generated coefficient graph.

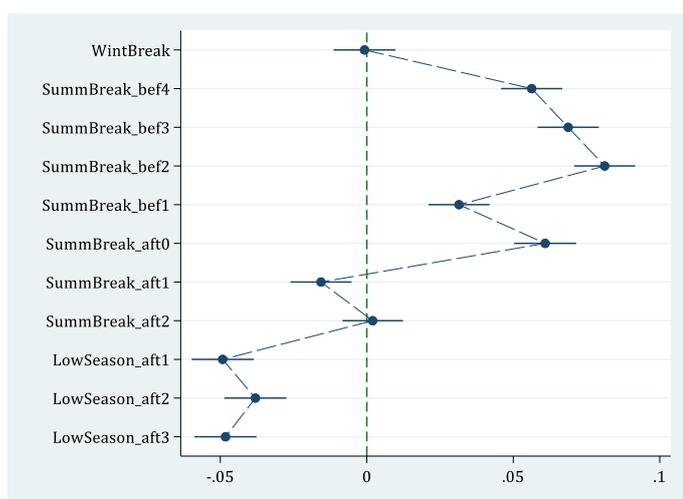

**Figure 1– Price Seasonality Coefficients Chart**

The results presented in Table 3 and Figure 1 support previously established findings: air ticket prices start to increase steadily around four months before the high season. They culminate in October (SummBreak_bef2), where we observe an approximate 8.1% price increase, holding other variables constant. Prices tend to either stabilize or exhibit slower growth beyond this point, a phenomenon also identified in Table 2. Notably, in November (SummBreak_bef1), the final month before summer, the rate of price growth decelerates, yielding the lowest coefficient in that period. The presence of three holidays in November could explain this, implying a shift in passenger demographics. A decrease in business travel and a potential increase in aircraft availability during holidays might apply downward pressure on price increases.

With summer's arrival, evidence suggests airlines introduce price promotions focused on the latter part of the year. Airlines aren't likely finalizing their full-year pricing strategies at this stage, as economic indicators remain uncertain until after Carnival and pending initial economic forecasts. As the low season begins, Table 3 shows negative and statistically significant LowSeason coefficients. This detailed seasonal analysis affirms earlier Table 2 observations, strengthening the evidence for consistent seasonal patterns in airline ticket pricing. Once we have studied the average seasonality within the sample period, we can proceed with our analysis of the pandemic event. In econometrics, an "event study" is an empirical method used to estimate the causal impact of a specific, significant event on a variable of interest. This relevant event can be any kind of exogenous change, such as a company merger, the implementation of a new policy, or a natural disaster. The core principle of an event study is to compare the behavior of the variable of interest before and after the event while accounting for other factors that might also influence the variable. When investigating an event, it's crucial to clearly define the period representing the specific occurrence under examination. In our case, aligning with the configuration of the Pandemic variable, we consider the pandemic's start to be February 2020 (when the first COVID-19 case was documented in Brazil). The end of the pandemic period is determined to be April 2022, coinciding with the Ministry of Health's declaration ending the Public Health Emergency of National Importance due to COVID-19. The intra-pandemic period is not included in our analysis.

Below is the code utilized to generate event study regressions and present the results. We conduct two regressions that capture the effect of granular seasonality.

```
*Before Pandemic
. reghdfe AirFare FuelPrice PaxDens MktConc LoadFactor
  Pandemic Trend WintBreak SummBreak_* LowSeason_* if
  YearMonth<=202001, absorb(CityPair)
. est store PrePandemic
*Post Pandemic
. reghdfe AirFare FuelPrice PaxDens MktConc LoadFactor
  Pandemic Trend WintBreak SummBreak_* LowSeason_* if
  YearMonth>202204, absorb(CityPair)
. est store PostPandemic
*show results table
. esttab PrePandemic PostPandemic, nocons nose not
  nogaps noobs b(%9.4f) varwidth(14) brackets
  aic(%9.0fc) bic(%9.0fc) ar2 scalar(N) sfmt(%9.0fc)
  addnote("Notes: Dependent variable - AirFare" "Fixed
  Effect estimation")
```

Table 4 presents the results of our regression analysis. Three significant findings emerge, aside from seasonality effects. First, the fuel coefficient (FuelPrice) is notably lower in the post-pandemic period than before the pandemic. This indicates that airlines are placing less emphasis on fuel price fluctuations, with competitive dynamics likely being the primary driver. Second, we observed a reduced price elasticity concerning aircraft occupancy rates (LoadFactor) in the post-pandemic era. This shift could stem from other cost factors exerting pressure on airline companies. Third, and highlighting these points, the estimated trend (Trend coefficient) has reversed. The pre-pandemic long-term trend suggested a price decline of roughly 2.5% every five years. However, this trend has changed direction in the post-pandemic period, signaling possible average price increases, ceteris paribus, of roughly 23% every five years. This strong upward trend is potentially unsustainable, reflecting the surge in demand associated with economic recovery, coupled with challenges in acquiring new aircraft to address that demand. The global travel market currently faces an aircraft shortage as economies worldwide rebound. This lack of aircraft restricts airlines' ability to expand capacity in line with demand, ultimately leading to pressure on prices. This shortage likely explains the observed reversal of the trend variable coefficient.

**Table 4– Event study: before and after the Pandemic**

```
                      (1)          (2)
                PrePandemic  PostPandemic

FuelPrice         0.1677***    0.0681*
PaxDens          -0.1608***   -0.1466***
MktConc           0.0889***   -0.0582***
LoadFactor        0.3020***    0.1889***
Trend            -0.0244***    0.2301***
WintBreak         0.0668***   -0.0293*
SummBreak_bef4    0.0822***    0.0688***
SummBreak_bef3    0.0878***    0.0876***
SummBreak_bef2    0.1178***    0.0559***
SummBreak_bef1    0.0686***    0.0313*
SummBreak_aft0    0.1107***    0.0546***
SummBreak_aft1    0.0263***   -0.0176
SummBreak_aft2    0.0181**    -0.0486**
LowSeason_aft1   -0.0265***   -0.1048***
LowSeason_aft2   -0.0125*     -0.0269*
LowSeason_aft3   -0.0138*     -0.0444**

adj. R-sq         0.647        0.722
AIC              -3,381       -2,558
BIC              -3,236       -2,439
N                37,441        8,373

Notes: Dependent variable - AirFare
Fixed Effect estimation
* p<0.05, ** p<0.01, *** p<0.001
```

Regarding seasonality, the results of the regressions were again displayed in the form of a coefficient plot, using the following command:

```
. coefplot PrePandemic PostPandemic, keep(WintBreak
  SummBreak_* SummBreak_* LowSeason_*) xline(0,
  lcolor(green) lpattern(dash)) scheme(s2color)
```

Figure 1 reveals the seasonality patterns of prices in Brazil's air transport sector before and after the pandemic. Post-pandemic (red), we observe a non-statistically significant price movement during the July holidays (WintBreak), contrasting with the pre-pandemic period (blue) where prices had a notable increase. Regarding the typical price increases leading into the high season (SummBreak_bef4) from August onwards, there were no significant differences between the two periods. However, a distinct and atypical pattern emerges in the post-pandemic context as peak season approaches. Flights likely reach capacity and off-season sales begin, leading to significantly more pronounced discounts. Estimated coefficients are generally smaller and negative, particularly at the end of the high season and the beginning of the low season. This potentially reflects a shift in passenger composition (leisure vs. business travel) post-pandemic compared to pre-pandemic. Airlines may be implementing more aggressively countercyclical pricing strategies with deeper discounts.

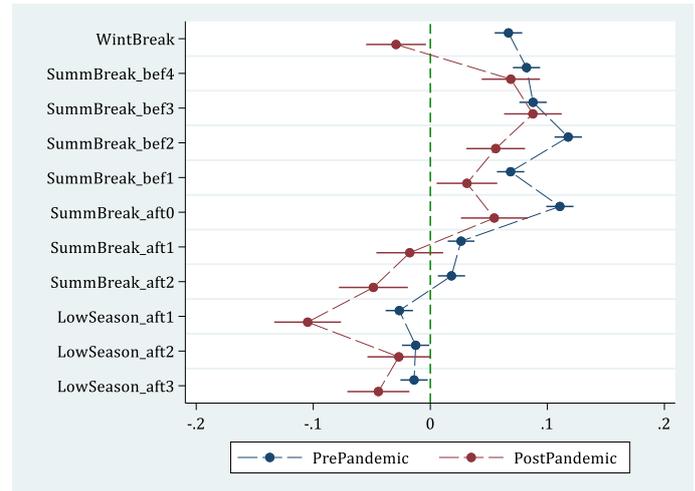

**Figure 2– Event Study Coefficients Chart**

In Figure 2, we observe that the increased discounts seen at the start of the low season begin to lessen as the season continues. This pattern suggests that in the final two months ('LowSeason_aft2' and 'LowSeason_aft3'), the difference between pre-pandemic and post-pandemic periods essentially disappears. This is likely because promotional seats fill more rapidly during this timeframe.

## V. CONCLUSIONS

This study developed an econometric model to examine the seasonality patterns of domestic air transport prices in Brazil, both before and after the COVID-19 pandemic. Our findings reveal statistically significant changes in these patterns. Notably, we observed a trend towards increased price discounts during the low season, alongside a long-term trajectory of substantial overall price increases. This intensified discrepancy between high and low season prices (seasonal deconcentration) could be attributed to the reduced seat capacity in the post-pandemic era, as well as the evolving composition of the domestic passenger market.

Understanding price seasonality is crucial for both businesses and consumers. It provides valuable insights into the patterns of supply and demand throughout the year, enabling the optimization of purchasing, selling, and marketing strategies. In light of this, our findings have implications for airline corporate policy. By incorporating them into revenue management practices, airlines can better accommodate varying seasonal demand stemming from both leisure and business travel. Aligning revenue management with consumer-focused market research could also facilitate strategies aimed at stabilizing the annual demand cycle through proactive counter-seasonal initiatives.

To support further research on this topic, we have made the data and code used in our econometric models available online. We encourage replications and the following potential extensions: 1. An event study comparing post-pandemic data from "treatment" routes (perhaps focusing on areas that experienced greater economic growth or lesser pandemic impact) with data from a less-affected "control" group. 2. Incorporation of multiple fixed effects (expanding the dimensions of the MFE Estimator) to examine regional interactions with seasonality. 3. Investigating the link between tourism-heavy routes/airports and seasonal concentration. 4. Exploring two-way fixed effects to consider idiosyncratic temporal factors and gain insights into the potential reversal of the long-term price trend identified here.

ACKNOWLEDGMENTS

The author wishes to thank the São Paulo State Research Support Foundation (FAPESP), grant no. 2020-06851, to the National Council for Scientific and Technological Development (CNPq), grant number 305439/2021-9. The author also thanks ITA colleagues Mauro Caetano, Marcelo Guterres, Evandro Silva, Giovanna Ronzani, Rogéria Arantes, Cláudio Jorge Pinto Alves, Mayara Murça, and Paulo Ivo Queiroz. Any errors are solely the responsibility of the author.

## REFERENCES

Alshuqaiqi, A., & Omar, S. I. (2019). Causes and implication of seasonality in tourism. Journal of Adv. Res. in Dynamical and Control Systems, 11(4), 1480-1486.
Boffa, F., & Succurro, M. (2012). The impact of search cost reduction on seasonality. Annals of Tourism Research, 39(2), 1176-1198.
Brito, I. R., Oliveira, A. V. M., & Dresner, M. E. (2021). An econometric study of the effects of airport privatization on airfares in Brazil. Transport Policy, 114, 338-349.
Cazanova, J., Ward, R. W., & Holland, S. (2014). Habit persistence in air passenger traffic destined for Florida. Journal of Travel Research, 53(5), 638-655.
Claussen, J., Essling, C., & Peukert, C. (2018). Demand variation, strategic flexibility and market entry: Evidence from the US airline industry. Strategic Management Journal, 39(11), 2877-2898.
Dobruszkes, F., Decroly, J. M., & Suau-Sanchez, P. (2022). The monthly rhythms of aviation: A global analysis of passenger air service seasonality. Transportation Research Interdisciplinary Perspectives, 14, 100582.
Garrigos-Simon, F. J., Narangajavana, Y., & Gil-Pechuan, I. (2010). Seasonality and price behaviour of airlines in the Alicante-London market. Journal of Air Transport Management, 16(6), 350-354.
Guven, M., Calik, E., Cetinguc, B., Guloglu, B., & Calisir, F. (2019). Assessing the effects of flight delays, distance, number of passengers and seasonality on revenue. Kybernetes, 48(9), 2138-2149.
Halpern, N. (2011). Measuring seasonal demand for Spanish airports: Implications for counter-seasonal strategies. Res. Transp. Bus. & Management, 1(1), 47-54.
Jann, B. (2014). Plotting regression coefficients and other estimates. The Stata Journal 14(4): 708-737.
Kraft, S., & Havlíková, D. (2016). Anytime? Anywhere? The seasonality of flight offers in Central Europe. Moravian Geographical Reports, 24(4), 26-37.
Law, R., Leung, R., Guillet, B. D., & Lee, H. A. (2011). Temporal changes of airfares toward fixed departure date. Journal of Travel & Tourism Marketing, 28(6), 615-628.
Merkert, R., & Webber, T. (2018). How to manage seasonality in service industries–the case of price and seat factor management in airlines. Journal of Air Transport Management, 72, 39-46.
Narangajavana, Y., Garrigos-Simon, F. J., García, J. S., & Forgas-Coll, S. (2014). Prices, prices and prices: A study in the airline sector. Tourism Mang, 41, 28-42.
Oliveira, A. V. M. (2024) AVDATA-BR: Uma base de dados aberta do transporte aéreo brasileiro. Communications in Airline Economics Research, n. 10011. Center for Airline Economics, S. J. Campos, Brazil.
Oliveira, A. V. M. (2023). Voar é para muitos. Os negócios das companhias aéreas e a popularização das viagens de avião no Brasil. Editora da PUCRS.
Perera, S., & Tan, D. (2019). In search of the "Right Price" for air travel: First steps towards estimating granular price-demand elasticity. Transportation Research Part A: Policy and Practice, 130, 557-569.
Póvoa, H., & Oliveira, A. V. M. (2013). Econometric analysis to estimate the impact of holidays on airfares. Journal of Transport Literature, 7, 284-296.
Resende, C. B., & Oliveira, A. V. M. (2023). Viagens de avião e inflação. In A. V. M. Oliveira (Org.), Voar é Para Muitos - Os Negócios das Companhias Aéreas e a Popularização das Viagens de Avião no Brasil (1ª ed., pp. 81-92). Porto Alegre: Editora Universitária da PUCRS (EDIPUCRS).
Reynolds-Feighan, A., Zou, L., & Yu, C. (2022). Seasonality in European and North American Air Transport Markets: Network Structures and Implications for Airline Performance and Recovery. Transportation Journal, 61(3), 284-304.
Ridderstaat, J., & Croes, R. (2020). A framework for classifying causal factors of tourism demand seasonality: An interseason and intraseason approach. Journal of Hospitality & Tourism Research, 44(5), 733-760.
Rosselló, J., & Sansó, A. (2017). Yearly, monthly and weekly seasonality of tourism demand: A decomposition analysis. Tourism Management, 60, 379-389.
Soysal, G. P., & Krishnamurthi, L. (2012). Demand dynamics in the seasonal goods industry: An empirical analysis. Marketing Science, 31(2), 293-316.
Turrión-Prats, J., & Duro, J. A. (2018). Tourist seasonality and the role of markets. Journal of Destination Marketing & Management, 8, 23-31.
Vergori, A. S., & Arima, S. (2022). Transport modes and tourism seasonality in Italy: By air or by road?. Tourism Economics, 28(3), 583-598.
Zou, L., Reynolds-Feighan, A., & Yu, C. (2022). Airline seasonality: An explorative analysis of major low-cost carriers in Europe and the United States. Journal of Air Transport Management, 105, 102272.
Zuidberg, J. (2017). Exploring the determinants for airport profitability: Traffic characteristics, low-cost carriers, seasonality and cost efficiency. Transportation Research Part A: Policy and Practice, 101, 61-72.

## OPEN CODES AND DATA

Correia, S. (2016). FTOOLS: Stata module to provide alternatives to common Stata commands optimized for large datasets. Statistical Software Components S458213, Boston College Department of Economics, revised 21 Aug 2023. Disponível em https://ideas.repec.org/c/boc/bocode/s458213.html.
Correia, S. (2017). REGHDFE: Stata module for linear and instrumental-variable/gmm regression absorbing multiple levels of fixed effects. Statistical Software Components S457874, Boston College Department of Economics. Disponível em: https://ideas.repec.org/c/boc/bocode/s457874.html
Jann, B. (2004). ESTOUT: Stata module to make regression tables. Statistical Software Components, Boston College Department of Economics, revised 12 Feb 2023. Available at https://ideas.repec.org/c/boc/bocode/s439301.html.
Jann, B. (2013). COEFPLOT: Stata module to plot regression coefficients and other results. Statistical Software Components, Boston College Department of Economics, revised 25 Feb 2023. Available at https://ideas.repec.org/c/boc/bocode/s457686.html.
Oliveira, A. V. M. (2024a). Airline price seasonality. Stata do-file. Available at https://github.com/avmoliveira/avdatabr/blob/main/caer10021_season.do.
Oliveira, A. V. M. (2024b). AVDATA-BR - Brazilian Aviation Data, Harvard Dataverse. Available at https://doi.org/10.7910/DVN/CRYXUZ.
Oliveira, A. V. M. (2024c). AVDATA-BR-CP - Brazilian Aviation Data - City-Pair Level, Harvard Dataverse. Available at https://dataverse.harvard.edu/api/access/datafile/avdatabr_cp_cae/8171526.
Wolfe, F. (2002). FSUM: Stata module to generate and format summary statistics. Statistical Software Components, Boston College Department of Economics, revised 06 May 2014. Available at https://ideas.repec.org/c/boc/bocode/s426501.html.